\documentclass[twocolumn]{article}
\usepackage{epsfig}

\usepackage{anysize}
\marginsize{2cm}{2cm}{2cm}{2cm}

\begin{document}

\title{Adaptive Dispersion Compensation for Remote Fiber Delivery of NIR Femtosecond Pulses} 

\author{S.H. Lee, A.L. Cavalieri$^*$, D.M. Fritz, M. Myaing$^{*}$ and D.A. Reis}

\date{\small FOCUS Center and Department of Physics, The University of
Michigan, Ann Arbor, Michigan 48109, USA \\
$^{*}$FOCUS Center and Applied Physics Program,The University of
Michigan, Ann Arbor, Michigan 48109, USA  \\ shl@umich.edu}


\maketitle

\abstract{We report on remote delivery of 25 pJ broadband near-infrared femtosecond light pulses from a Ti:sapphire laser through 150
meters of single-mode optical fiber.  Pulse distortion due to dispersion is overcome with pre-compensation using adaptive pulse shaping
techniques, while nonlinearities are mitigated using an SF10 rod for the final stage of pulse compression.  Near transform limited pulse
duration of 130 fs is measured after the final compression.}

\section*{Introduction}
                                                                                                                                                                                                         
Delivery of ultrafast laser pulses over long distances has important applications for accelerator diagnostics and pump-probe experiments performed at future XFEL's and next generation light
sources\cite{Emma,Wilke,LCLS}. Due to mechanical constraints and stability concerns, transmission of light through an optical fiber is a flexible alternative to free space propagation. However,
large material dispersion and nonlinear effects inside the fiber core introduce technical challenges. Sub-picosecond pulse transmission has been achieved over lengths in excess of a kilometer.
These optical links operate at telecommunication wavelengths near the zero dispersion point (1.55 um) of standard single mode fiber using specialty dispersion compensating fiber
(DCF)\cite{Pelusi00} Recently, Chang \emph{et al.}\cite{Chang98} used a spatial light modulator (SLM) in addition to the DCF to correct for higher order dispersion.  For Ti:Sapphire lasers
operating at 800 nm,the effects of dispersion are much more severe and no DCF is readily available.  Nonlinear response of the fiber ultimately limits the pulse energy and duration for these
techniques.  In particular, self phase modulation can lead to spectral compression for negatively chirped pulses used in pre-compensation\cite{Planas93}.  With proper dispersion compensation, the
result of this nonlinearity is at best a longer, albeit transform limited, pulse duration \cite{Washburn00,Myaing00}. Clark \emph{et. al.}\cite{Clark01} demonstrated group velocity dispersion
compensated 100 fs, 0.5 nJ pulses at 800 nm upon propagation through $\sim$1 m of fiber using a combination of temporal and spectral compression.  Their technique is scalable to longer distances
with additional elements for the compensation of higher order dispersion but relies on the fiber nonlinearity; consequently, it is sensitive to the characteristics of the input pulse.

In this letter, we demonstrate a novel technique for the transport of femtosecond Ti:Sapphire laser pulses through 150 meters of standard single mode polarization preserving optical fiber using
adaptive pulse shaping.  We pre-compensate for the large group velocity and higher order dispersion of the fiber so the fiber transport serves to compress the pre-chirped input pulse.  At 800 nm
wavelength, material dispersion of the fused-silica core is expected to be the dominant source of dispersion\cite{Agrawal}.  Both group velocity dispersion (GVD) and third order dispserion (TOD)
are significant for long distance propagation (360 fs$^2$/cm and 280 fs$^3$/cm respectively).  However, adaptive pulse shaping makes precise \emph{a priori} knowledge of the dispersion irrelevant
in our system.  Relatively high pulse energies can be accommodated while maintaining linear pulse propagation by avoiding full compression in the optical fiber.  Final pulse compression is
performed external to the fiber, where the intensities are low, using a high dispersion glass rod.

In our experiment, sub-50 fs optical pulses are generated by a Kerr-lens mode-locked Ti:Sapphire laser at a central wavelength of 800 nm and a 102MHz repetition rate.  The transport system
comprises a series of dispersive elements designed to achieve zero net dispersion at the output(see Figure 1). The first element is a 1200 line/mm grating pair in a parallel
geometry\cite{Treacy}.  The grating pair provides the negative GVD to compensate for the positive GVD (normal dispersion) imparted by the combination of the fiber and SF10 rod (approximately
$-5.4 \times10^6$ fs$^2$, or 16 ps/nm) but introduces TOD with the same sign and comparable magnitude to the fiber. The second element is an \emph{arbitrary} pulse shaper that is used to correct
for uncompensated dispersion (TOD and higher, as well as any residual GVD from misalignment of the grating pair).  The third element is 150 m of single mode polarization preserving optical
fiber\cite{theFiber}, used for pulse delivery and the majority of the compression.  Finally, the 25 cm long SF10 glass rod at the exit of the fiber compresses the pulses further completing the
transport system.  A $\sim$1.5\% transmission efficiency was obtained through the system yielding a pulse energy of 25 pJ for a 1.5nJ pulse input, limited primarily by the grating efficiencies
and non-optimal coupling into the fiber.  

  \begin{figure}[htbp]
  \centering
  \includegraphics[width=8.3cm]{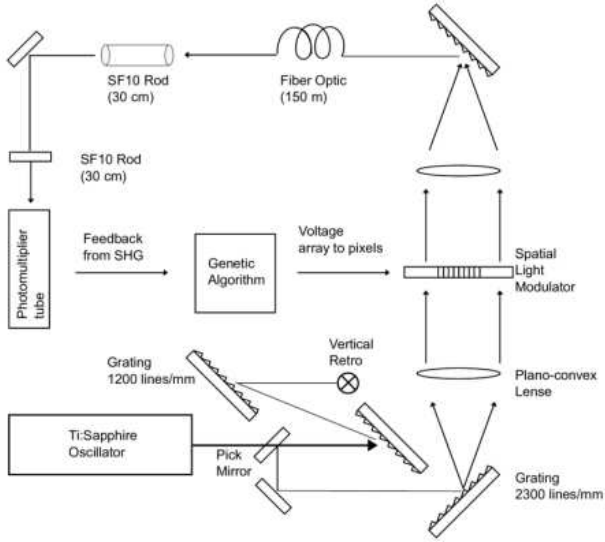}
  \caption{Experimental Setup.  A femtosecond laser pulse is sent through a sequence of dispersion compensation elements before
entering150 meters of single mode optical fiber.}
  \end{figure}

The order of the dispersive elements in our transport system are critical since ultrafast pulses easily reach intensities where contributions from nonlinear phenomena become significant, due to
the small core diameter of the fiber optic (5um)\cite{Agrawal}.  Nonlinear effects in the fiber transport are minimized by placing the GVD compensation and the adaptive pulse shaping scheme
before the fiber input to pre-stretch the pulse.  Here, the peak laser intensity is at its lowest value, but, increases as the material dispersion of the fiber re-compresses the optical pulse.
The pulses exit the fiber slightly chirped ($\sim 0.1$ ps/nm) to keep the peak intensity $< \sim1$GW/cm$^2$; the final compression is performed in the SF10 after the transverse beam size is
expanded to 5mm.  An additional advantage of this pre-compensation scheme is that spatial chirp, introduced by the pulse shaping, is filtered by the fiber.

The pulse shaper contains a liquid crystal mask spatial light modulator (SLM) consisting of rectangular pixels whose indices of refraction depend on the voltage applied across them.  Since the
SLM is located in the Fourier plane of a \emph{4f} stretcher, the transmitted temporal pulse shape is altered by applying phase shifts to the individual Fourier components of the pulse. Figure 2
shows a cross-correlation of the transmitted pulse when the pulse shaper is not active.  The TOD introduced by both the optical fiber and gratings is clearly significant.  To compensate for
residual dispersion in the transport system, we use a genetic algorithm (GA)\cite{Baumert97} to search for the optimum voltage combination applied across the mask The strength of using this
search algorithm is that it does not require any calculation of dispersion.  Instead, voltages across the pixels are stored in an array known as an ``individual", and the initial population of
individuals is generated at random.  Transmitted pulses are frequency doubled in a BBO crystal. The second harmonic efficiency generation is used as a feedback paramter for the genetic algorithm,
and based on this fitness factor, a set of individuals is selected and used for the generation of a new population for evaluation.  The feedback is valid so long as the conversion efficiency in
the BBO is proportional to the intensity, and thus inversely proportional to the pulse duration.  The genetic algorithm proceeds until a plateau is reached in the fitness.  A solution is normally
found after 300 iterations ($\sim$30 minutes).

  \begin{figure}[htbp]
  \centering
  \includegraphics[width=8.3cm]{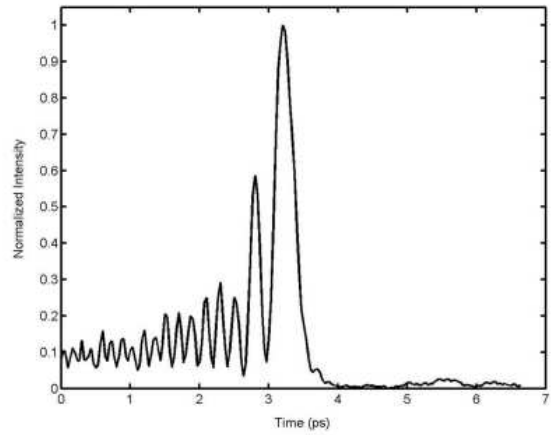}
  \caption{Cross-correlation between input pulse and output pulse from the optical fiber after the group velocity dispersion
compensation. FWHM is limited to 600fs and effect of third order dispersion is evident.}
  \end{figure}

Figure 3 shows autocorrelation traces of the the pulses transmitted through the fiber and SF10 before and after running the GA.  After the GA, the pulses have been compressed to 130 fs assuming a
near Gaussian profile. This result corresponds to better than 1.4 times the diffraction limit for a Gaussian pulse at a bandwidth of 10 nm.  Further compression of the output pulse is hindered by
the limited bandwidth coupled into the fiber (see Figure 4).  Currently the bandwidth is limited by clipping at the input of the fiber. We estimate that sub-100fs pulses should be achievable with
the current system.  Past this limit, as the bandwidth is increased linearly to support a shorter pulse, the resolution of the SLM must be increased cubically.  This occurs since the pulse shaper
compensates primarily for 3rd order dispersion in the transport system.  The need for increased resolution in the SLM can be avoided by using a fixed phase mask, so that the adaptive pulse
shaping makes only small corrections to the pulse shape.  With these improvements sub-50 fs, nJ scale pulse delivery should be possible.  The system has been successfully put to use at the
Subpicosecond Pulse Source (SPPS) experiment at the Stanford Linear Accelerator Center where it is used for a relative timing diagnostic for ultrafast x-ray diffraction experiments
\cite{AdrianInPrep}.

  \begin{figure}[htbp]
  \centering
  \includegraphics[width=8.3cm]{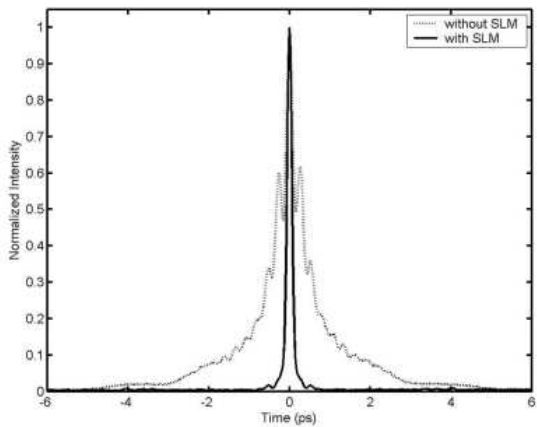}
  \caption{Auto-correlation of output pulse from the fiber delivery system.  With SLM on, FWHM is reduced to 130 fs.}
  \end{figure}

  \begin{figure}[htbp]
  \centering
  \includegraphics[width=8.3cm]{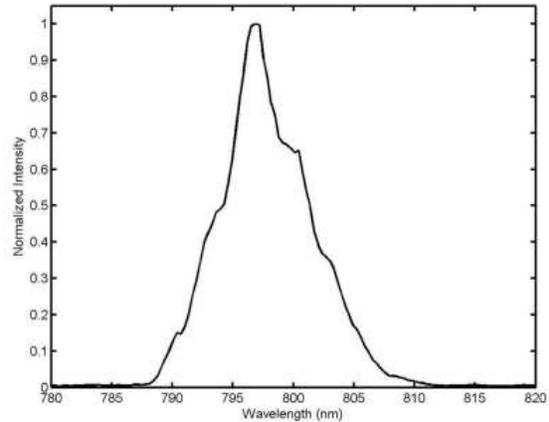}
  \caption{Spectrum of the output pulse. Spatial chirp from mirrors and grating pair is cleaned out by the fiber.}
  \end{figure}

\section*{Acknowledgments}
                                                                                                                                                                                                         
The authors thank B. Pearson, M. DeCamp, and P. Bucksbaum for valuable discussions and the SPPS collaboration for use of the laser oscillator.  This research was supported in part by the US
Department of Energy, under contract no. DE-FG02-99ER45643 and the National Science Foundation, FOCUS Physics Frontier Center under grant no. 011436.

\end{document}